%% file: iclr2027_conference.tex
\documentclass{article} 
\usepackage{iclr2027_conference,times}

\input{math_commands.tex}

\usepackage{hyperref}
\usepackage{url}
\usepackage{graphicx} 
\usepackage{amsmath}
\RequirePackage{algorithm}
\RequirePackage{algorithmic}
\usepackage{xspace}

\usepackage[utf8]{inputenc} 
\usepackage[T1]{fontenc}    
\usepackage{hyperref}       
\usepackage{url}            
\usepackage{booktabs}       
\usepackage{amsfonts}       
\usepackage{nicefrac}       
\usepackage{microtype}      
\usepackage{xspace}
\usepackage{amsmath,amssymb,amsthm}
\usepackage{mathtools} 
\usepackage{adjustbox}
\usepackage{titletoc}
\usepackage[dvipsnames]{xcolor}
\usepackage{tcolorbox}                

\theoremstyle{definition}

\usepackage{multirow}

\newcommand*{\method}{CurvFM\xspace}
\newcommand*{\tx}{Transition1x\xspace}
\newcommand*{\swap}{Transition1x-2p3p4p\xspace}
\newcommand*{\tmc}{Transition1x-TMC\xspace}

\title{Curvature-Aware Flow Matching for\\ Molecular Structure Generation}

\author{%
Samir Darouich$^{1,2,3}$, Juliane Geisler$^{2}$, Jacob W.~Toney$^{1}$, Tanja Bien$^{2}$,\\
\textbf{Johannes K{\"a}stner$^{3}$, Heather J.~Kulik$^{1,4}$, Mathias Niepert$^{2}$\thanks{Corresponding author.}}\\[4pt]
$^{1}$Dept.\ of Chemical Engineering, MIT, USA\\
$^{2}$Institute for Artificial Intelligence, University of Stuttgart, Germany\\
$^{3}$Institute for Theoretical Chemistry, University of Stuttgart, Germany\\
$^{4}$Dept.\ of Chemistry, MIT, USA\\[2pt]
\texttt{samir.darouich@ki.uni-stuttgart.de}, \texttt{juliane.geisler@web.de},\\
\texttt{jwt@mit.edu}, \texttt{tanja.bien@ki.uni-stuttgart.de}, \\
\texttt{kaestner@theochem.uni-stuttgart.de}, \texttt{hjkulik@mit.edu},\\
\texttt{mathias.niepert@ki.uni-stuttgart.de}%
}

\iclrfinalcopy 
\begin{document}

\maketitle
\lhead{Preprint}

\input{00_abstract}
\input{01_introduction}
\input{02_related_work}
\input{03_background}
\input{04_method}

\input{05_experiments}
\input{06_discussion}

\subsection*{AI use statement}

In this work, we used generative AI tools to aid and polish writing, to draft sections of the paper, and to support research execution through code generation for analysis and evaluation scripts. We have not used generative AI tools to generate synthetic datasets, develop theoretical models or conceptual frameworks, formulate mathematical claims, propose or refine hypotheses, design or provide feedback on research methodology or experiments, implement the methods presented in this work, support qualitative or thematic data analysis, or interpret results. All AI-assisted work was reviewed by the authors, with AI-assisted text revised as needed and AI-generated code checked for correctness. The experimental design, model development, and interpretation of results were conducted by the authors. We take responsibility for the final content of this work, including all text, claims, and artifacts produced with the aid of generative AI.




\subsection*{Reproducibility statement}

All components needed to reproduce our results are described in the paper. The residual and the full training objective are defined in Sec.~\ref{sec:method}, with a single training step summarized in Algorithm~\ref{alg:method}. The datasets, splits, and evaluation metrics are specified in Sec.~\ref{sec:experiments}, and the metric definitions are given in full in Appendix~\ref{sec:metrics}. Appendix~\ref{sec:implementation} reports the training setup for both tasks, including the checkpoints we start from, the number of epochs, learning rates, batch sizes, hardware, and the procedure used to select the residual weight and the time-weighting exponent. Additionally, it describes how the reference Hessians for \swap{} and \tmc{} were computed and how the Hessian prediction model was finetuned on them, together with its accuracy on each dataset. 


\subsubsection*{Acknowledgments}

This research was funded by the Ministry of Science, Research and the Arts Baden-Wuerttemberg in the Artificial Intelligence Software Academy (AISA) and by the Deutsche Forschungsgemeinschaft (DFG, German Research Foundation) -- Project number 569019417. We also acknowledge the support of the Stuttgart Center for Simulation Science (SimTech) and thank the International Max Planck Research School for Intelligent Systems (IMPRS-IS) for support. J.W.T. was partially supported by a Leslye Miller Fraser and Darryl M. Fraser Fellowship from the MIT School of Engineering. T.B. was supported by the Deutsche Forschungsgemeinschaft (DFG, German Research Foundation) under Germany’s Excellence Strategy – EXC 2120/1 – 390831618. H.J.K. is supported by a Simon Family Faculty Research Innovation Fund and an Alfred P. Sloan Fellowship in Chemistry. 

\clearpage

\bibliography{literature}
\bibliographystyle{iclr2027_conference}

\newpage

\begin{center}
    \hrule 
    \startcontents[sections]\vbox{\vspace{4mm}\sc \LARGE Curvature-Aware Flow Matching for\\ Molecular Structure Generation \\\sc\small \textbf{Additional Material}} \vspace{5mm} \hrule height .5pt
    \printcontents[sections]{l}{0}{\setcounter{tocdepth}{2}}
\end{center}

\newpage

\appendix
\input{07_appendix}

\end{document}

%% file: math_commands.tex
\usepackage{amsmath,amsfonts,bm}

\def\eqref#1{equation~\ref{#1}}

\def\1{\bm{1}}

\DeclareMathAlphabet{\mathsfit}{\encodingdefault}{\sfdefault}{m}{sl}
\SetMathAlphabet{\mathsfit}{bold}{\encodingdefault}{\sfdefault}{bx}{n}



%% file: 00_abstract.tex
\begin{abstract}
    Generative models produce three-dimensional molecular structures with high geometric accuracy, yet they are trained on distributions of atomic coordinates without explicit access to the underlying physics. The potential energy surface (PES) describes how the energy changes around a molecular geometry, and the root mean square deviation (RMSD) between a generated structure and its reference captures this local landscape only indirectly. Supervising the PES during training could close this gap, but conformers and transition states (TSs) are stationary points at which the forces vanish, so first-order quantities carry little information about them. What remains is the local curvature, encoded in the Hessian of the PES. Quantum-chemical Hessians are too expensive to be available at the scale required for training, but recently proposed machine-learned predictors make curvature inexpensive enough to supervise directly. We introduce \method{}, a flow matching (FM) formulation whose training objective constrains the curvature at the generated structure. A frozen Hessian prediction model is evaluated at the generated endpoint and at the reference structure, which removes the need for quantum-chemical Hessians and keeps the residual cheap enough to impose at every training step. Across TS and conformer generation, \method{} substantially improves physical fidelity at unchanged global structural accuracy, reducing the median deviation in maximum force by $63\%$ on GEOM-QM9 and raising the fraction of valid TSs by up to $10\%$ across three reaction datasets.
\end{abstract}

%% file: 01_introduction.tex
\section{Introduction}

Generative models have become the dominant approach for producing three-dimensional molecular structures of small organic molecules, covering conformer ensembles as well as reaction transition states (TSs)~\citep{jing2022torsional,hassan2024flow,kim2024diffusion,duan2025optimal}. They are trained on coordinate-based objectives and evaluated with geometric metrics such as the root mean square deviation (RMSD), which measures how far a generated structure sits from its reference. Agreement in this sense is only loosely related to agreement in energy, even though the geometry determines the energy exactly. A generated molecule can sit close to the reference under RMSD and still contain a steric clash or a distorted bonding environment that raises its energy considerably~\citep{galustian2025goflow,darouich2026adaptive,zhao2025harnessing}. Such errors are costly because barrier heights and Boltzmann-weighted ensemble properties depend exponentially on the energy, so even moderate energetic errors translate into large errors in the predicted property.

\begin{figure}[t]
\centering
\includegraphics[width=0.98\textwidth]{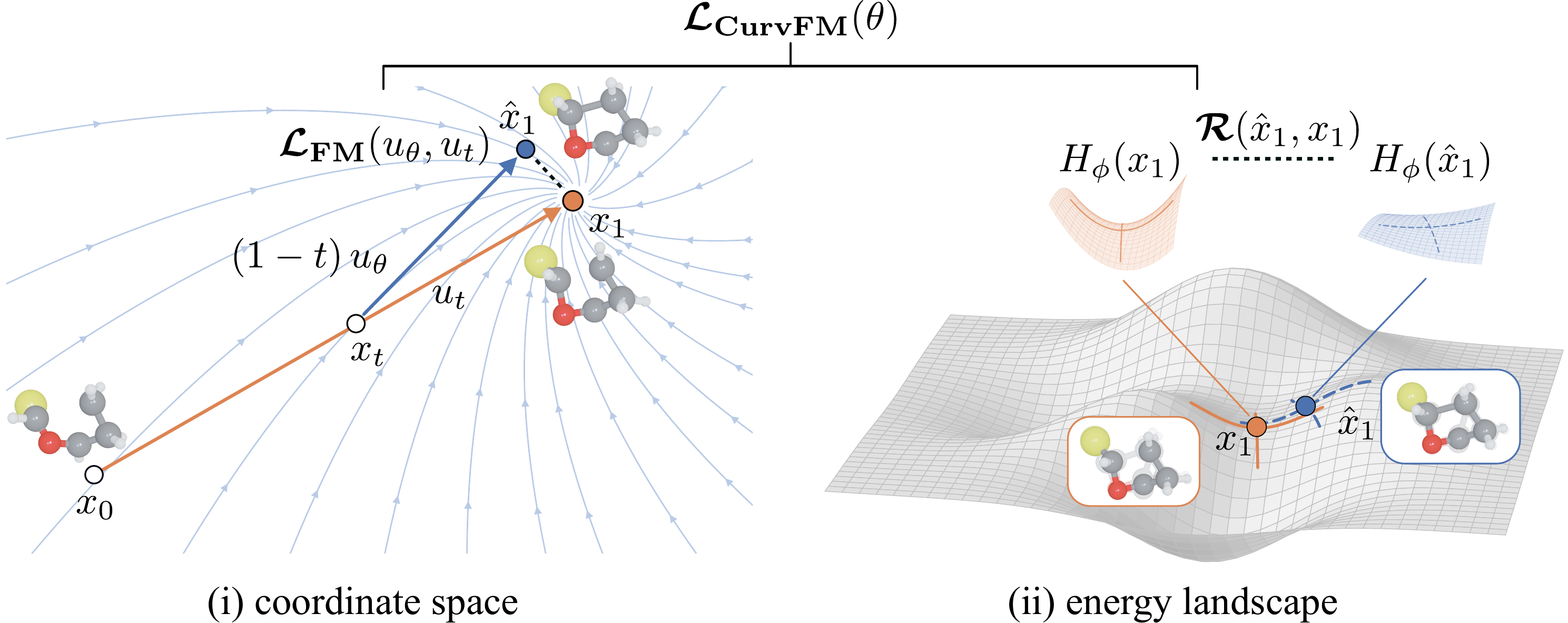}
\caption{
\textbf{Conceptual illustration of \method}.
\textbf{(i)} At an interpolated sample $x_t$, the model predicts a velocity $u_\theta(x_t, t)$ that is regressed against the conditional velocity field $u_t$ pointing to the target structure $x_1$. Integrating the velocity field to $t=1$ gives an estimate $\hat{x}_1$ of the clean endpoint.
\textbf{(ii)} Both $\hat{x}_1$ and $x_1$ are passed through a frozen Hessian predictor $H_\phi$, and the residual $\mathcal{R}(\hat{x}_1,x_1)$ compares the curvature of the PES at the two geometries. Penalizing their disagreement encourages the generated structure to lie at the right place on the energy landscape, which the FM loss alone does not account for.
}
\label{fig:method_overview}
\end{figure}

The relation between geometry and energy is given by the potential energy surface (PES). What makes a generated structure physically meaningful is where it sits on this surface, not how close it lies to a reference geometry. Supervising a model on both, the geometry and its position on the PES, is therefore the natural way to close this gap. Constraining a generative model against a known physical law during training is an established idea, developed most systematically for partial differential equations (PDEs), where the governing equation is available in closed form and its residual can be evaluated on the generated sample at every step~\citep{bastek2025physicsinformed,baldan2026physics}. For molecules no such closed-form operator exists, and evaluating the PES at quantum-chemical accuracy inside a training loop is prohibitively expensive. Physical information has therefore been introduced through surrogates, in the form of physics-based potentials or learned energy models that steer the trajectory at inference~\citep{wohlwend2025boltz,xu2025energy}, and of force-field gradients or rewards used to finetune a pretrained model~\citep{berlaga2025flowback,zhou2025guiding}.

All of these signals are of zeroth or first order in the energy, which limits what they can express. The structures most commonly targeted by molecular generative models, such as conformers and TSs, are stationary points of the PES. A conformer is a local minimum and a TS is a first-order saddle point, so the forces vanish at either and no first-order signal separates them. What distinguishes the two is the local curvature, encoded in the Hessian of the energy with respect to the atomic positions, which is therefore the lowest-order quantity that characterizes these structures at all. Supervising a generative model on curvature has until recently been impractical. Machine-learned interatomic potentials (MLIPs) give access to the PES at a fraction of the quantum-chemical cost. Obtaining a Hessian from one, however, requires differentiating the model twice. Inside a training loop the residual must in addition be differentiated to update the generator, which adds considerable overhead to every step. Models that predict the Hessian directly rather than by differentiation avoid this entirely, returning the full matrix from a single forward pass~\citep{cui2025large,burger2025shoot}. Curvature thereby becomes affordable as a training signal.

We introduce \method{}, a \textbf{curv}ature-aware \textbf{f}low \textbf{m}atching (FM) formulation that supervises the curvature of the PES at the structures the model generates. During training, the predicted velocity is used to form an estimate of the clean endpoint, which is passed through a frozen Hessian prediction model and compared against the prediction of the same model at the ground-truth endpoint. Evaluating both sides with the same surrogate cancels its systematic error and removes the need for reference Hessians, so the residual can be applied to datasets for which no such labels exist. Across TS and conformer generation, \method{} leaves global agreement with the reference geometry unchanged while placing the generated structures closer to the stationary points they are meant to occupy. The median deviation in maximum force decreases by $63\%$ on GEOM-QM9, and the fraction of valid TS rises up to 10\% across three reaction datasets. Our main contributions are as follows.

\begin{itemize}
    \item We introduce \method{}, a second-order supervision scheme for generative models that constrains the curvature of the PES at the generated sample using a frozen Hessian predictor.
    \item We show that a surrogate-based residual must be evaluated with the same model on both sides, so that its systematic error cancels and no quantum-chemical labels are required.
    \item Across four datasets and two architectures, \method{} improves physical fidelity at unchanged global structural accuracy, and the resulting structures require fewer steps to relax to a true stationary point.
\end{itemize}

%% file: 02_related_work.tex
\section{Related work}

\paragraph{Physics-constrained generative modeling.}

Physical constraints have been incorporated into generative models most extensively for PDEs, either at inference, by projecting or guiding samples toward the constraint set during integration~\citep{huang2024diffusionpde,jacobsen2025cocogen,utkarsh2026physics}, or during training, by adding a residual term to the generative objective~\citep{bastek2025physicsinformed,baldan2026physics}. For molecules, where no such residual can be evaluated in closed form, the physical signal comes from a surrogate. Most operate at inference, steering the sampling trajectory with physics-based potentials or with a learned energy model to remove clashes and other unphysical artifacts~\citep{wohlwend2025boltz,xu2025energy,mark2025feynman}. Others act after training, finetuning a pretrained model against energy and force rewards derived from a machine-learned force field~\citep{zhou2025guiding,li2026elign} or steering its vector field with force-field gradients via adjoint matching~\citep{berlaga2025flowback}.

\paragraph{Molecular structure generation.}

Conformer and TS generation are both instances of generating 3D geometry conditioned on a graph-level specification. For conformers, the condition is a single molecular graph and the target is a set of geometries compatible with it. Distance-geometry and rule-based generators such as RDKit~\citep{riniker2015better} remain widely used, while diffusion and flow matching models have emerged as a data-driven alternative~\citep{luo2021predicting,shi2021learning,xu2022geodiff,jing2022torsional,liu2025nextmol,cao2025efficient,xu2025energy,darouich2026symdrift}. For TSs, the condition is a reactant and product pair and the target is the saddle-point geometry separating them. Locating such a point classically means running an iterative or path-based search~\citep{jonsson1998nudged,henkelman_climbing_2000,banerjee_search_1985,henkelman_dimer_1999}, whose cost is dominated by the electronic-structure calls it requires. Replacing these calls with a MLIP reduces the cost by orders of magnitude while retaining the guarantees of an explicit optimization~\citep{yuan2024analytical,wander_cattsunami_2024}. The search remains iterative, however, and its outcome depends on how well the potential describes the reactive region of the surface, which is where reference data is typically scarcest. Diffusion and flow models instead sidestep the search altogether by generating the saddle-point geometry directly~\citep{duan2023accurate,kim2024diffusion,duan2025optimal,galustian2025goflow,darouich2026adaptive,darouich2026robust}.

%% file: 03_background.tex
\section{Background}
\label{sec:background}

\paragraph{Physics-based flow matching.}

Flow matching learns a time-dependent velocity field $u_\theta$ that transports a tractable source distribution $p_0$ to the data distribution $p_1$~\citep{lipman2023flow}. Training is done in a simulation-free fashion by drawing a noise sample $x_0$ and a data sample $x_1$ and interpolating between them. Under the optimal-transport formulation~\citep{tong2024improving} the interpolant is the straight line $x_t = (1-t)\,x_0 + t\,x_1$, along which the conditional velocity is constant, $u_t = x_1 - x_0$. The network is then regressed against this target,
\begin{equation}
  \mathcal{L}_{\mathrm{FM}}(u_\theta, u_t) = \mathbb{E}_{t, x_t} \left\| u_\theta(x_t, t) - u_t \right\|_2 ,
\end{equation}
and sampling integrates the learned field from $t=0$ to $t=1$ with a numerical ODE solver.

Physical constraints can be added to this objective through a residual $\mathcal{R}$ that penalizes violations of a known physical law. Such a residual cannot be evaluated on $x_t$, which interpolates between noise and data and carries no physical meaning, so it is imposed instead on an estimate $\hat{x}_1$ of the clean endpoint obtained from the predicted velocity, giving $\mathcal{L} = \mathcal{L}_{\mathrm{FM}} + \| \mathcal{R}(\hat{x}_1) \|_2$~\citep{bastek2025physicsinformed, baldan2026physics}. This substitution introduces Jensen's gap~\citep{zhang2025physics}, since for a nonlinear residual $\mathbb{E}[\mathcal{R}(x_1) \mid x_t] \neq \mathcal{R}(\mathbb{E}[x_1 \mid x_t])$. The quantity minimized at intermediate noise levels is therefore not necessarily the residual of interest. PBFM~\citep{baldan2026physics} reduces this discrepancy by unrolling the ODE over $n$ steps of size $(1-t)/n$ before evaluating $\mathcal{R}$, and by weighting the residual loss with $t^p$ to down-weight the imprecise endpoint estimates obtained near $t=0$.

\paragraph{The potential energy surface and its curvature.}
The PES relates the energy of a molecular system to its geometric configuration and underlies the theoretical description of chemical reactivity~\citep{lewars2024computational}. Quantum-chemical methods such as density functional theory (DFT) provide approximate PESs, but exploring one exhaustively is infeasible for all but the smallest systems, since a nonlinear molecule with $N$ atoms spans $3N-6$ internal degrees of freedom. In practice, one is rarely interested in the surface as a whole, but in the few points that carry chemical meaning and in the local behavior of the surface around them~\citep{dewyer2018methods}. Conformers and TSs are two such points, both stationary and therefore characterized by the local curvature of the surface. That curvature is encoded in the Hessian $H \in \mathbb{R}^{3N \times 3N}$, the second derivative of the energy with respect to the atomic positions. Its eigenvectors are the normal modes of the system and the corresponding eigenvalues give the curvature along them. Six eigenvalues vanish, corresponding to the global translations and rotations that leave the energy unchanged. A conformer is a local minimum, so the remaining $3N-6$ eigenvalues are positive, whereas a TS is a first-order saddle point, so exactly one is negative and its eigenvector spans the reaction coordinate. Obtaining Hessians at the quantum-chemical level is expensive, which motivates the use of surrogate models. MLIPs replace the quantum-chemical PES with a surrogate fitted to energy and force labels, and are differentiable, so Hessians can in principle be obtained by automatic differentiation and used in second-order TS optimization schemes~\citep{yuan2024analytical}. Curvature is, however, not recovered reliably from first-order labels alone~\citep{zhao2025harnessing,cuarare2025global,bheemaguli2026evaluation}, which has motivated Hessian supervision during MLIP training~\citep{rodriguez2025does,cui2025large,rodriguez2026projected,yin2026hessian}. A more direct route bypasses differentiation altogether and predicts the Hessian as a model output~\citep{wu2026machine}, as in Hessian Interatomic Potentials (HIP)~\citep{burger2025shoot}, which read the full matrix out of an SE(3)-equivariant backbone.

%% file: 04_method.tex
\section{Curvature-Aware Flow Matching}
\label{sec:method}

Purely geometric supervision leaves an important aspect of molecular structure unconstrained. We characterize this gap below and then introduce \method{}, which closes it by supervising the model on curvature information.

\subsection{The limits of geometric supervision}

Coordinate-based objectives and the RMSD-type metrics used to evaluate them treat all displacement directions alike, but the PES does not. A displacement along a soft torsion changes the energy negligibly, whereas one of the same magnitude along a bond stretch changes it by orders of magnitude more. Averaging squared displacements over all $3N$ coordinates therefore obscures exactly the directions in which the energy varies most, and does so increasingly as the molecule grows. The consequences are documented across both tasks. TS energies are highly sensitive to small structural perturbations~\citep{duan2025optimal}, and generated conformers must generally be relaxed before ensemble properties can be reliably computed~\citep{jing2022torsional}. A structure can therefore lie close to the reference and still be far from the stationary point it is meant to represent~\citep{zhao2025harnessing,galustian2025goflow,darouich2026adaptive,darouich2026robust,de2026toward}.

\subsection{Curvature as a training signal}

Closing the gap between geometric agreement and physical validity requires a training signal that refers to the PES itself, rather than to the reference geometry alone. Evaluating the local shape of the surface at the structures the model generates makes their position on the PES accessible during training. Since conformers and TSs are both stationary points, at which the forces vanish, this information must come from the curvature, and the Hessian is the lowest-order object in which a minimum and a first-order saddle point remain distinguishable.

What limits the use of this signal is the cost of obtaining it during training. Quantum-chemical Hessians from DFT are far too expensive to evaluate for every sample in every batch. MLIPs provide a cheaper surrogate, but obtaining a Hessian from one requires differentiating the model twice with respect to the atomic positions. This costs $3N$ backward passes per structure and produces an activation footprint that grows with the square of the system size. In a generative training loop the burden compounds further, since the residual must itself be differentiated with respect to the generator parameters, which places the Hessian evaluation on the backward path of every update and exposes third derivatives of the surrogate. Models that predict the Hessian directly avoid both difficulties, obtaining the full matrix in a single forward pass~\citep{burger2025shoot}. Curvature then costs no more than an energy evaluation, and the backward path through the frozen surrogate remains first order, so the residual can be imposed at every training step at marginal cost in memory and runtime.

\subsection{Residual formulation}

Curvature enters the residual $\mathcal{R}$ through a direct Hessian predictor $H_\phi : \mathbb{R}^{3N \times 3} \to \mathbb{R}^{3N \times 3N}$, which remains frozen throughout training and is queried at the structures the generative model produces. The residual compares the curvature at the generated endpoint $\hat{x}_1$ against the curvature at the corresponding reference structure $x_1$,
\begin{equation}
  \mathcal{R}(\hat{x}_1, x_1) = \mathcal{D}\bigl( H_\phi(\hat{x}_1),\, H_\phi(x_1) \bigr),
  \label{eq:general_residual}
\end{equation}
where $\hat{x}_1$ is an estimate of the clean endpoint obtained during training and $\mathcal{D}$ is a discrepancy between two curvature matrices. The choice of reference curvature, either the surrogate prediction or a quantum-chemical calculation at $x_1$, has a substantial effect on the resulting training signal. Using the surrogate on both sides means the two predictions carry largely the same systematic error, which cancels in the comparison, as we verify in Appendix~\ref{sec:additional_exp_ts}. The residual then reflects the quality of $\hat{x}_1$ rather than the inaccuracies of the surrogate itself. Avoiding quantum-chemical reference Hessians also makes the residual applicable to datasets for which no such labels exist. The reference term depends solely on the data, so in practice $H_\phi(x_1)$ is precomputed once for the training set and a training step requires only a single surrogate evaluation.

The endpoint $\hat{x}_1$ at which the residual is evaluated is obtained from the velocity predicted by the FM model. At time $t$, an intermediate sample $x_t$ is drawn along the transport path and the model predicts a velocity $u_\theta(x_t, t)$. Integrating this velocity to $t=1$ yields an estimate of the clean endpoint,
\begin{equation}
  \hat{x}_1 = x_t + (1-t)\, u_\theta(x_t, t),
  \label{eq:endpoint}
\end{equation}
which we refer to as the single-step case. Unrolling the ODE over $n$ steps of size $(1-t)/n$ brings $\hat{x}_1$ closer to the true endpoint at the cost of an additional model call per step.

The choice of $\mathcal{D}$ determines which aspect of the curvature is constrained. The most direct instantiation compares the two Hessian matrices entrywise,
\begin{equation}
  \mathcal{D}_{\mathrm{MAE}} = \frac{1}{(3N)^2} \sum_{i,j} \left| \bigl[H_\phi(\hat{x}_1)\bigr]_{ij} - \bigl[H_\phi(x_1)\bigr]_{ij} \right| ,
  \label{eq:residual_mae}
\end{equation}
which supervises the full curvature at the generated geometry without singling out any part of the spectrum. An alternative restricts the comparison to the modes that define the stationary point. Writing $V$ and $\Lambda$ for the $k$ lowest eigenvectors and eigenvalues of $H_\phi(x_1)$, this variant penalizes the deviation of the predicted Hessian projected onto that subspace,
\begin{equation}
  \mathcal{D}_{\mathrm{sub}} = \frac{1}{k^2} \sum_{i,j} \left| \bigl[ V^\top H_\phi(\hat{x}_1) V - \Lambda \bigr]_{ij} \right| ,
  \label{eq:residual_sub}
\end{equation}
following the projection used for curvature supervision of interatomic potentials~\citep{burger2025shoot}. Diagonalizing the reference side rather than the prediction keeps the eigendecomposition outside the computational graph, which avoids the numerical instabilities that arise when differentiating through eigenvectors of a matrix with near-degenerate eigenvalues. Both comparisons are entrywise and therefore require that the two Hessians refer to a common frame, since the Hessian is equivariant and transforms blockwise as $H_{ij} \mapsto R H_{ij} R^\top$ under a rotation $R$. Aligning $x_0$ to $x_1$ before constructing the transport path is sufficient, as the entire path and the equivariant surrogate then share the frame of $x_1$. Unless stated otherwise we use $\mathcal{D}_{\mathrm{MAE}}$, and analyze $\mathcal{D}_{\mathrm{sub}}$ and the combination of both in Sec.~\ref{sec:additional_experiments}.

The full objective weights the residual by $t^p$ to down-weight the imprecise endpoint estimates obtained near $t=0$,
\begin{equation}
  \mathcal{L}_\text{\method}(\theta) = \mathcal{L}_{\mathrm{FM}}(u_\theta, u_t) + \lambda\, t^p\, \mathcal{R}(\hat{x}_1, x_1),
  \label{eq:total_loss}
\end{equation}
and gradients propagate to $\theta$ through $\hat{x}_1$ only. The two terms act on complementary aspects of the same sample, with the velocity loss constraining the geometry and the residual constraining the curvature at the geometry the model produces. Figure~\ref{fig:method_overview} illustrates the method conceptually, and Algorithm~\ref{alg:method} summarizes a single training step for the $\mathcal{D}_{\mathrm{MAE}}$ residual.

\begin{algorithm}[ht]
\caption{One training step of \method{} using the $\mathcal{D}_{\mathrm{MAE}}$ residual.}
\label{alg:method}
\begin{algorithmic}[1]
\REQUIRE Flow model $u_\theta$, frozen Hessian model $H_\phi$, source sample $x_0$, target sample $x_1$, residual weight $\lambda$, exponent $p$
\STATE $x_0 \leftarrow \mathrm{align}(x_0, x_1)$ \hfill \COMMENT{common frame for the whole path}
\STATE $t \sim \mathcal{U}(0,1)$
\STATE $x_t \leftarrow (1-t)\,x_0 + t\,x_1$
\STATE $v \leftarrow u_\theta(x_t, t)$
\STATE $\mathcal{L}_{\mathrm{FM}} \leftarrow \| v - (x_1 - x_0) \|_2^2$
\STATE $\hat{x}_1 \leftarrow x_t + (1-t)\,v$ \hfill \COMMENT{single-step endpoint estimate}
\STATE $H_{\mathrm{pred}} \leftarrow H_\phi(\hat{x}_1)$
\STATE $H_{\mathrm{ref}} \leftarrow \mathrm{sg}\!\left[ H_\phi(x_1) \right]$ \hfill \COMMENT{target side, no gradient}
\STATE $\mathcal{R} \leftarrow \tfrac{1}{(3N)^2} \sum_{i,j} \bigl| [H_{\mathrm{pred}}]_{ij} - [H_{\mathrm{ref}}]_{ij} \bigr|$
\STATE $\mathcal{L} \leftarrow \mathcal{L}_{\mathrm{FM}} + \lambda\, t^p\, \mathcal{R}$
\RETURN $\mathcal{L}$
\end{algorithmic}
\end{algorithm}

%% file: 05_experiments.tex
\section{Experiments}
\label{sec:experiments}

We evaluate \method on two complementary molecular structure generation tasks, TS generation and molecular conformer generation. Across both settings, we investigate whether curvature supervision improves the physical fidelity of generated structures while preserving their geometric accuracy, and analyze which aspects of the local energy landscape account for the observed gains.

\subsection{Transition state generation}

\paragraph{Datasets and Metrics.}

As a first task, we consider TS generation and evaluate \method{} on three datasets of increasing chemical diversity. \tx{} contains 10,073 gas-phase reactions at the $\omega$B97x/6-31G(d) level involving H, C, N, and O with up to seven heavy atoms~\citep{schreiner2022transition1x}. Building on its reactions, \swap{} applies elemental swaps up to the 4p group to give 14,000 reactions, and \tmc{} comprises 36,446 reactions covering ten catalytically relevant TMs with up to 26 atoms, both at the GFN2-xTB level~\citep{darouich2026robust}. Following~\citet{darouich2026robust}, each dataset is split randomly into training, validation, and test sets with an 80/10/10 ratio. As Hessian surrogate we use HIP~\citep{burger2025shoot}. For \swap{} and \tmc{} we compute reference Hessians for the reactant, TS, and product structures at the level of theory of the respective dataset, and finetune the public checkpoint on them to cover the elements they contain. Further details are given in Appendix~\ref{sec:training_hip}. Structural accuracy is measured by the RMSD for global similarity and the distance matrix absolute error (DMAE) for finer structural differences. Since the central claim of this work concerns physical rather than geometric fidelity, we evaluate the generated structures directly on the PES, computing energies and forces at the corresponding level of theory. We then assess the curvature that \method{} is trained on by comparing the reference Hessian of the generated structure against that of the true TS, computed at the same level of theory rather than with the surrogate.

\paragraph{Results.}

To isolate the effect of curvature supervision, we compare against the identical model trained with the structural FM loss alone under the same training budget. Table~\ref{tab:structural_metrics_summary} reports both variants within the React-OT~\citep{duan2025optimal} framework across all three datasets. The global structural metrics remain essentially unchanged, consistent with their weak coupling to energetic accuracy discussed above. Resolving the error by internal coordinate shows where the geometry does change (Table~\ref{tab:bond_metrics_summary}). On \swap{} and \tmc{}, the median bond and angle errors decrease by up to $17$ and $8\%$, while the softer torsions and non-bonded distances remain unchanged. The residual thus acts on the degrees of freedom along which the energy varies most steeply, and the energetic metrics improve accordingly. Both the median barrier error and the maximum force decrease by about $21\%$ on these two datasets, reducing the median maximum force from $33.50$ to $26.43$ and from $46.74$ to $36.58$~kcal/mol/\AA{}. On \tx{}, where the baseline is already strong, the gain is roughly $3\%$. These gains are obtained at an increase in training time of $30\%$ and in peak memory of $12\%$, while sampling remains unchanged (Appendix~\ref{sec:cost}).

\begin{table}[ht]
\centering
\caption{Structural and energetic metrics for TS generation. Best in bold, second-best underlined.}
\label{tab:structural_metrics_summary}
\begin{adjustbox}{max width=\linewidth}
\begin{tabular}{llcccccccc}
\toprule
 &  & \multicolumn{2}{c}{RMSD (\AA) $\downarrow$} & \multicolumn{2}{c}{D-MAE (\AA) $\downarrow$} & \multicolumn{2}{c}{$\Delta E$ (kcal/mol) $\downarrow$} & \multicolumn{2}{c}{$F_\text{max}$ (kcal/mol/\AA) $\downarrow$} \\
\cmidrule(lr){3-4} \cmidrule(lr){5-6} \cmidrule(lr){7-8} \cmidrule(lr){9-10}
Dataset & Model & Mean & Median & Mean & Median & Mean & Median & Mean & Median \\
\midrule
\tx & React-OT & 0.188 & 0.108 & 0.068 & 0.042 & \textbf{3.89} & \underline{1.35} & \underline{33.99} & \underline{26.07} \\
\tx & React-OT$_\text{\method}$ & 0.188 & 0.108 & 0.068 & 0.042 & \underline{4.06} & \textbf{1.31} & \textbf{33.50} & \textbf{25.33} \\
\midrule
\swap & React-OT & \textbf{0.196} & \underline{0.134} & \underline{0.083} & \underline{0.060} & \underline{5.85} & \underline{2.53} & \underline{40.96} & \underline{33.50} \\
\swap & React-OT$_\text{\method}$ & \underline{0.197} & \textbf{0.133} & \textbf{0.082} & \textbf{0.058} & \textbf{5.19} & \textbf{2.00} & \textbf{32.69} & \textbf{26.43} \\
\midrule
\tmc & React-OT  & 0.294 & \textbf{0.231} & \underline{0.125} & \underline{0.101} & \underline{9.92} & \underline{6.40} & \underline{55.78} & \underline{46.74} \\
\tmc & React-OT$_\text{\method}$ & 0.294 & \underline{0.232} & \textbf{0.124} & \textbf{0.100} & \textbf{8.27} & \textbf{4.91} & \textbf{43.08} & \textbf{36.58} \\
\bottomrule
\end{tabular}
\end{adjustbox}
\end{table}

Table~\ref{tab:hessian_metrics_summary} evaluates the curvature directly, comparing the Hessians of the generated and the reference structures, both computed with quantum chemistry at the level of theory of the respective dataset. The fraction of valid TSs, those with exactly one negative eigenvalue, increases on all three datasets, from $70$ to $74\%$ on \tx{}, from $57$ to $67\%$ on \swap{}, and from $32$ to $41\%$ on \tmc{}. On \swap{} and \tmc{} these gains are reflected in a more accurate curvature overall, with both error measures improving and the median eigenvalue error decreasing by $21$ and $19\%$ respectively, whereas on \tx{} the full Hessian error remains unchanged.

To understand where these gains originate, we resolve the Hessian into its eigenmodes, each describing a collective displacement of the atoms together with the curvature of the energy along it. The reactive mode is the single direction of negative curvature along which the reaction proceeds, while the remaining vibrational modes range from soft bends and torsions to stiff bond stretches. As shown in Table~\ref{tab:hessian_metrics_summary_detailed}, the structural objective alone already recovers the reactive mode well, with a median overlap of $95.9\%$ on \swap{} between the predicted direction and its reference, measured as the absolute cosine similarity between the two eigenvectors. Curvature supervision raises this only to $96.6\%$. The residual therefore does not primarily sharpen the reaction coordinate, but corrects the curvature of the surrounding energy landscape. 

This raises the question of whether the gain requires curvature at all, or whether any physical signal evaluated at the endpoint would suffice. Forces are the natural first-order alternative. They are cheaper to obtain, but they vanish at stationary points and so carry little information near the structures being generated. Replacing the Hessian with the forces at the same two endpoints, with the loss weight and time weighting retuned, recovers only about a third of the gain. Despite supervising the maximum force directly, the force residual reduces it by less than half as much as the curvature residual does (Table~\ref{tab:structural_metrics_summary_ablation_force}).

Given that second-order information is needed, the remaining question is which part of the spectrum carries it. Table~\ref{tab:hessian_metrics_summary_ablation_loss} compares the full residual of Eq.~\ref{eq:residual_mae} against the subspace variant of Eq.~\ref{eq:residual_sub}, which constrains only the projection onto the $k=8$ lowest reference eigenvectors and eigenvalues, and against the combination of both. The subspace variant stays close to the purely structural baseline, reducing the median eigenvalue error by $6\%$ and leaving the valid TS rate essentially unchanged, in contrast to the $21\%$ and ten point improvements obtained with the full residual. The gap is widest among the stiff modes, where the median relative error decreases by $7\%$ under the subspace variant and by $27\%$ under the full residual, since a residual confined to the low-lying part of the spectrum leaves the stiff curvature, which dominates the Hessian, unconstrained. Adding the subspace term to the full residual degrades every metric slightly, so the targeted supervision does not merely fail to help but competes with the denser signal it is combined with.

\begin{table}[ht]
\centering
\caption{Hessian metrics for TS generation. Hessians of both the generated and the ground-truth structures are computed with quantum chemistry at the level of theory of the respective dataset. Best in bold, second-best underlined.}
\label{tab:hessian_metrics_summary}
\begin{adjustbox}{max width=\linewidth}
\begin{tabular}{llccccc}
\toprule
 &  & \multicolumn{2}{c}{Hessian MAE (kcal/mol/\AA$^2$) $\downarrow$} & \multicolumn{2}{c}{Eigenvalue MAE (kcal/mol/\AA$^2$) $\downarrow$} & \multicolumn{1}{c}{Valid TS (\%) $\uparrow$} \\
\cmidrule(lr){3-4} \cmidrule(lr){5-6} \cmidrule(lr){7-7}
Dataset & Model & Mean & Median & Mean & Median & Mean \\
\midrule
\tx & React-OT & \textbf{13.43} & \underline{7.95} & \textbf{25.74} & \underline{17.69} & \underline{70.00} \\
\tx & React-OT$_{\text{\method}}$ & \underline{13.52} & \textbf{7.94} & \underline{25.99} & \textbf{16.93} & \textbf{74.00} \\
\midrule
\swap & React-OT & \underline{12.13} & \underline{8.40} & \underline{30.89} & \underline{22.95} & \underline{57.00} \\
\swap & React-OT$_{\text{\method}}$ & \textbf{11.39} & \textbf{7.52} & \textbf{25.31} & \textbf{18.08} & \textbf{67.00} \\
\midrule
\tmc & React-OT & \underline{11.13} & \underline{9.14} & \underline{33.24} & \underline{27.57} & \underline{32.00} \\
\tmc & React-OT$_{\text{\method}}$ & \textbf{10.32} & \textbf{8.32} & \textbf{26.76} & \textbf{22.43} & \textbf{41.00} \\
\bottomrule
\end{tabular}
\end{adjustbox}
\end{table}

\begin{figure}[ht]
\centering
\includegraphics[width=\textwidth]{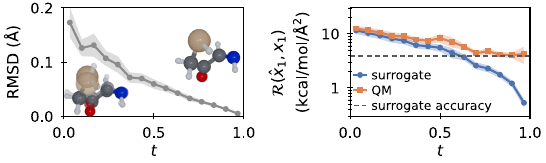}
\caption{\textbf{A quantum-chemical reference bounds the resolvable residual.} Quality of the one-shot prediction $\hat{x}_1$ on \swap{} as a function of the FM time $t$. Markers joined by lines are binned means and shaded regions are confidence intervals of the mean. \textbf{Left:} Kabsch-aligned RMSD between $\hat{x}_1$ and the reference TS $x_1$, which depends only on the predicted geometry and is therefore identical for both variants. \textbf{Right:} the residual $\mathcal{R}(\hat{x}_1, x_1)$ for the same predictions, evaluated against the surrogate and against the QM Hessian of the reference structure. The dashed line marks the error of the surrogate itself at the reference geometry.}
\label{fig:residual_source}
\end{figure}

The source of the reference curvature matters as much as its formulation. A quantum-chemical target might appear preferable to a surrogate one, since it removes an approximation from the comparison, but it degrades the model below the purely structural baseline (Table~\ref{tab:structural_metrics_summary_ablation_hessian}). The reason is an irreducible floor of the surrogate model. Comparing the surrogate prediction at $\hat{x}_1$ against a quantum-chemical Hessian at $x_1$ leaves a residual that does not vanish even when the two geometries coincide. Once the endpoint estimate becomes accurate enough that its own contribution falls below this floor, the residual stops varying with $\hat{x}_1$ and the gradient carries no usable information. Figure~\ref{fig:residual_source} shows this directly. The RMSD of the endpoint estimate decreases with $t$ for both variants, but the residual evaluated against a quantum-chemical target stops decreasing once it reaches the surrogate accuracy. The surrogate residual continues below it, since most of its error is shared between the two geometries and cancels (Appendix~\ref{sec:additional_exp_ts}).

Training against a surrogate raises the question of whether the model inherits its bias. If it did, the gains would appear larger when evaluated with the surrogate than with quantum chemistry. Repeating the Hessian evaluation with HIP instead gives nearly the same picture (Table~\ref{tab:hessian_metrics_summary_hip}), so the gains reflect improved curvature rather than agreement with the surrogate used during training.

\subsection{Conformer generation}

\paragraph{Datasets and Metrics.}

As a second task, we consider conformer generation on GEOM-QM9~\citep{axelrod2022geom}, comprising 133,258 small molecules whose reference ensembles were obtained with CREST~\citep{pracht2024crest}. We use the splits of~\citet{ganea2021geomol}, with $80\%$ of the molecules for training, $10\%$ for validation, and a held-out set of 1,000 molecular graphs for testing. For a molecule with $K$ reference conformers we generate $2K$ samples, capped at 32, and apply the chirality correction of~\citet{ganea2021geomol}. Structural accuracy is measured by coverage (COV) and average minimum RMSD (AMR), both reported as recall and precision. Recall quantifies how completely the generated ensemble reproduces the reference conformers and precision how many of the generated conformers are themselves accurate, with a conformer counted as covered when it falls within $\delta = 0.5$~\AA{}. As in the TS setting, these metrics compare geometries and say nothing about where the structures sit on the potential energy surface. We therefore relax all conformers with GFN2-xTB~\citep{bannwarth2019gfn2} and compare Boltzmann-weighted ensemble properties between the generated and reference ensembles, namely the energy $E$, the minimum energy $E_\text{min}$, the dipole moment $\mu$, and the HOMO-LUMO gap $\Delta \epsilon$. Finally, we assess the curvature directly by computing Hessians for up to 32 conformers per molecule and testing whether the generated structures are local minima, that is, whether their spectra are free of negative eigenvalues.

\paragraph{Results.}

To allow a controlled comparison, we take the publicly available ET-Flow checkpoint and finetune it for 50 further epochs, once with the structural FM loss alone and once with the \method{} objective. The global structural metrics remain essentially unchanged (Table~\ref{tab:etflow_structural_metrics_summary}), whereas the energetic metrics improve substantially under residual training (Table~\ref{tab:etflow_ensemble_metrics_summary}). For the generated structures, the median deviation in maximum force decreases by $63\%$, from $15.91$ to $5.81$~kcal/mol/\AA{}, and the deviation in minimum energy falls from $0.42$ to $0.12$~kcal/mol. After relaxation with GFN2-xTB, both models converge to nearly identical ensembles, confirming that each recovers the intended conformational distribution. What separates them is how close the generated structures already are to it. That proximity is a matter of curvature, and Table~\ref{tab:etflow_hessian_metrics_summary} shows how it changes. The median fraction of generated conformers that are local minima rises to $100\%$, so a majority of molecules now yield fully valid ensembles, while the mean is marginally lower. Training on curvature therefore places the generated structures nearer to genuine minima. This is reflected in the relaxation itself, where they reach equilibrium in fewer optimization steps, with the median dropping from $5.4$ to $4.8$ (Table~\ref{tab:etflow_optimization_metrics_summary}).

\method{} is model-agnostic, since the residual acts on the endpoint estimate rather than on the architecture that produces it. This raises the question of how its effect compares to that of a stronger architecture. We therefore compare ET-Flow with \method{} against DiTMC~\citep{frank2025sampling}, an independently developed model that reaches better structural accuracy on GEOM-QM9 (Table~\ref{tab:etflow_structural_metrics_summary_ablation_ditmc}). The two reach comparable physical fidelity, with DiTMC ahead on the ensemble energies and ET-Flow with \method{} ahead on the maximum force (Table~\ref{tab:etflow_ensemble_metrics_summary_ablation_ditmc}). Curvature supervision applied to the weaker model therefore closes the gap in physical fidelity to the stronger architecture, and remains available to it.

\begin{table}[ht]
\centering
\caption{Molecule conformer generation results on GEOM-QM9 ($\delta$ = 0.5\AA). Best in bold, second-best underlined.}
\label{tab:etflow_structural_metrics_summary}
\begin{adjustbox}{max width=\linewidth}
\begin{tabular}{lcccccccc}
\toprule
 & \multicolumn{2}{c}{Coverage-R (\%) $\uparrow$} & \multicolumn{2}{c}{AMR-R (\AA) $\downarrow$} & \multicolumn{2}{c}{Coverage-P (\%) $\uparrow$} & \multicolumn{2}{c}{AMR-P (\AA) $\downarrow$} \\
\cmidrule(lr){2-3} \cmidrule(lr){4-5} \cmidrule(lr){6-7} \cmidrule(lr){8-9}
Model & Mean & Median & Mean & Median & Mean & Median & Mean & Median \\
\midrule
ET-Flow & \underline{94.95} & 100.00 & \underline{0.088} & \underline{0.037} & \textbf{91.45} & 100.00 & \textbf{0.120} & \underline{0.055} \\
ET-Flow$_\text{\method}$  & \textbf{95.97} & 100.00 & \textbf{0.080} & \textbf{0.035} & \underline{91.33} & 100.00 & \underline{0.121} & \textbf{0.054} \\
\bottomrule
\end{tabular}
\end{adjustbox}
\end{table}

\begin{table}[ht]
\centering
\caption{Ensemble property deviations on GEOM-QM9 computed at GFN2-xTB, before and after geometry optimization. Best in bold, second-best underlined.}
\label{tab:etflow_ensemble_metrics_summary}
\begin{adjustbox}{max width=\linewidth}
\begin{tabular}{llcccccccccc}
\toprule
 &  & \multicolumn{2}{c}{$|\Delta \langle E \rangle|$ (kcal/mol)} $\downarrow$ & \multicolumn{2}{c}{$|\Delta \langle F_\text{max} \rangle|$ (kcal/mol/Å)} $\downarrow$ & \multicolumn{2}{c}{$|\Delta \langle \mu \rangle|$ (D)} $\downarrow$ & \multicolumn{2}{c}{$|\Delta \langle \Delta \epsilon \rangle|$ (kcal/mol)} $\downarrow$ & \multicolumn{2}{c}{$|\Delta \langle E_{min} \rangle| $ (kcal/mol)} $\downarrow$ \\
\cmidrule(lr){3-4} \cmidrule(lr){5-6} \cmidrule(lr){7-8} \cmidrule(lr){9-10} \cmidrule(lr){11-12}
Model & Stage & Mean & Median & Mean & Median & Mean & Median & Mean & Median & Mean & Median \\
\midrule
ET-Flow & Generated & \underline{0.64} & \underline{0.49} & \underline{16.90} & \underline{15.91} & \underline{0.16} & \underline{0.05} & \underline{1.78} & \underline{1.18} & \underline{0.58} & \underline{0.42} \\
ET-Flow$_\text{\method}$ & Generated & \textbf{0.33} & \textbf{0.15} & \textbf{6.37} & \textbf{5.81} & \textbf{0.15} & \textbf{0.04} & \textbf{1.05} & \textbf{0.44} & \textbf{0.31} & \textbf{0.12} \\
\midrule
ET-Flow & Optimized & \textbf{0.23} & \textbf{0.05} & 0.77 & 0.78 & \underline{0.15} & \textbf{0.03} & \underline{0.72} & \underline{0.21} & \textbf{0.18} & 0.02 \\
ET-Flow$_\text{\method}$  & Optimized & \underline{0.27} & \underline{0.06} & 0.77 & 0.78 & \textbf{0.14} & \underline{0.04} & \textbf{0.65} & \textbf{0.20} & \underline{0.23} & 0.02 \\
\bottomrule
\end{tabular}
\end{adjustbox}
\end{table}

%% file: 06_discussion.tex
\section{Discussion}

\paragraph{Limitations.}
\method{} requires a Hessian prediction model for the chemical domain of interest, which currently restricts it to systems covered by the available surrogates. Reference Hessians for training such a model remain expensive to compute, so extending the method to domains with no suitable predictor requires either transferring an existing model out of distribution or generating new reference data. Evaluating the residual adds one surrogate call per training step increasing training time and memory accordingly, although sampling remains unaffected. The residual is also only as informative as the surrogate is smooth. Since both sides are evaluated with the same model, systematic error cancels, but the gradient still relies on the surrogate responding consistently to changes in geometry, which is not guaranteed far from its training distribution. Finally, the gains we observe are consistent but do not place the generated structures at genuine stationary points, so a subsequent relaxation remains necessary before they can be used.

\paragraph{Future work.}

The Hessian carries more information than the classification of a stationary point. Its spectrum determines a range of observable second-order properties, so the same residual could supervise any of them and enable generation conditioned on such a target, for instance a vibrational spectrum. Another direction concerns the surrogate itself. Since both evaluations of the residual are performed with the same model, its systematic error cancels, and the accuracy the surrogate needs to reach is an open question. Establishing this would clarify whether cheaper or more narrowly trained Hessian models can be used without loss of quality.

%% file: 07_appendix.tex
\section{Implementation details}
\label{sec:implementation}

\subsection{Training details}

\paragraph{Hessian prediction model.}
\label{sec:training_hip}
For \tx{} we use the publicly available EquiformerV2 HIP checkpoint~\citep{burger2025shoot} without modification, since it is trained at the same level of theory and on the same elements. For \swap{} and \tmc{} we compute reference Hessians for every reactant, TS, and product structure by finite differences of the analytical forces, using ASE~\citep{ase_paper} together with the Python interface of GFN2-xTB~\citep{bannwarth2019gfn2}. The checkpoint is then finetuned on these Hessians for 400 epochs with a batch size of 5 and a cosine learning rate schedule from $5 \times 10^{-4}$ to $5 \times 10^{-5}$. The accuracy of the resulting surrogates is verified on the whole dataset, giving a Hessian MAE of 0.54, 3.51, and 3.25~kcal/mol/\AA$^2$ on \tx{}, \swap{}, and \tmc{}. For GEOM-QM9 we reuse the \swap{} model, which shares the GFN2-xTB level of theory, and confirm its accuracy on a subset of 5000 GEOM-QM9 conformers, which include Fluorine, a element not present in the \tx or \swap dataset, still leading to a similar Hessian MAE of 3.37~kcal/mol/\AA$^2$.

\paragraph{Transition state generation.}

We start from the adapted OA-ReactDiff checkpoint of~\citet{darouich2026robust}, which replaces the one-hot atom-type encoding of the original LEFTNet backbone~\citep{du2023new} with a learnable embedding of the atomic number supplemented by selected electronic properties, so that elements beyond H, C, N, and O are covered. On each dataset, the checkpoint is first finetuned for 200 epochs with the React-OT flow matching loss at a learning rate of $2.5 \times 10^{-4}$. From this shared starting point, both variants are trained for a further 200 epochs at the same learning rate, the baseline with the FM loss alone and \method{} with the residual of Eq.~\ref{eq:residual_mae}. The weight $\lambda$ and the exponent $p$ are optimized on the validation set using Bayesian optimization. Training uses a batch size of 16 on a single A100 GPU.

\paragraph{Molecular conformer generation.}

We follow ET-Flow~\citep{hassan2024flow} and use a harmonic prior defined by the connectivity of the molecular graph, removing the center of mass from each sampled configuration to enforce translational invariance. Starting from the publicly available GEOM-QM9 checkpoint, both variants are finetuned for a further 50 epochs, the baseline with the FM loss alone and \method{} with the residual of Eq.~\ref{eq:residual_mae}. The weight $\lambda$ and the exponent $p$ are again optimized on the validation set using Bayesian optimization. Training uses a fixed learning rate of $2.5 \times 10^{-4}$ and a batch size of 64 on a single A100 GPU.

\subsection{Evaluation Metrics}
\label{sec:metrics}

\paragraph{Transition state generation.}

Structural agreement between a generated structure $\hat{\mathbf{x}}_1$ and its reference $\mathbf{x}_1$ is quantified after Kabsch alignment by
\begin{equation}\label{eq:rmsd_definition}
    \text{RMSD}(\hat{\mathbf{x}}_1,\mathbf{x}_1) = \left( \frac{1}{N} \sum_{i=1}^{N} \big\| \hat{\mathbf{x}}_{1,i} - \mathbf{x}_{1,i} \big\|^2 \right)^{1/2},
\end{equation}
where $N$ is the number of atoms and $\mathbf{x}_{1,i} \in \mathbb{R}^3$ collects the Cartesian coordinates of atom $i$. Since the RMSD summarizes the deviation of the structure as a whole, we additionally report the distance matrix mean absolute error (DMAE), which is sensitive to local distortions and requires no alignment,
\begin{equation}\label{eq:dmae_definition}
    \text{DMAE}(\hat{\mathbf{x}}_1,\mathbf{x}_1) = \frac{1}{N(N-1)} \sum_{i \neq j} \big| \hat{d}_{ij} - d_{ij} \big|,
\end{equation}
with $d_{ij} = \| \mathbf{x}_{1,i} - \mathbf{x}_{1,j} \|$ the interatomic distance between atoms $i$ and $j$.

Physical fidelity is assessed on the PES. The deviation in barrier height follows from the electronic energies of the two structures,
\begin{equation}\label{eq:delta_e_definition}
    \Delta E = \big| E(\hat{\mathbf{x}}_1) - E(\mathbf{x}_1) \big| ,
\end{equation}
which coincides with the error in the barrier of the corresponding reaction, since both energies are referenced to the same reactant. How close a generated structure lies to a stationary point is measured by the largest force component acting on it,
\begin{equation}\label{eq:f_max_definition}
    F_\text{max} = \max_{i,j} \big| f_{ij}(\hat{\mathbf{x}}_1) \big| ,
\end{equation}
with $f_{ij}$ the component $j \in \{x,y,z\}$ of the force on atom $i$. In contrast to the preceding metrics, $F_\text{max}$ is evaluated on the generated structure alone and vanishes exactly at a stationary point.

Curvature is evaluated by comparing the Hessian $H(\hat{\mathbf{x}}_1) \in \mathbb{R}^{3N \times 3N}$ of the generated structure against the Hessian $H(\mathbf{x}_1)$ of the reference TS, both computed at the level of theory of the respective dataset. Since the Hessian is equivariant, the generated structure is rotated into the frame of the reference before the comparison,
\begin{equation}\label{eq:hessian_mae_definition}
    \text{Hessian MAE} = \frac{1}{(3N)^2} \sum_{i,j} \big| H_{ij}(\hat{\mathbf{x}}_1) - H_{ij}(\mathbf{x}_1) \big| .
\end{equation}
Writing $\lambda_1 \leq \dots \leq \lambda_{3N}$ for the eigenvalues of $H(\mathbf{x}_1)$ and $\hat{\lambda}_1 \leq \dots \leq \hat{\lambda}_{3N}$ for those of $H(\hat{\mathbf{x}}_1)$, the corresponding spectral error is
\begin{equation}\label{eq:eigenvalue_mae_definition}
    \text{Eigenvalue MAE} = \frac{1}{3N} \sum_{i=1}^{3N} \big| \hat{\lambda}_i - \lambda_i \big| ,
\end{equation}
The eigenvalue MAE treats all modes alike, although their physical roles and magnitudes differ considerably. We therefore resolve the spectrum further, working in mass-weighted coordinates and projecting out translations and rotations with the projector built from the respective geometry, which leaves $3N-6$ vibrational modes ordered by their eigenvalues $\omega_1 \leq \dots \leq \omega_{3N-6}$. The lowest of these is the reactive mode, and its direction is compared through the absolute cosine similarity of the corresponding eigenvectors,
\begin{equation}\label{eq:overlap_definition}
    \text{Neg. Mode Overlap} = \big| \hat{\mathbf{v}}_1 \cdot \mathbf{v}_1 \big| ,
\end{equation}
where the absolute value accounts for eigenvectors being defined only up to sign. 

The remaining modes are grouped by their reference wavenumber, with those below $500$~cm$^{-1}$ classified as soft, corresponding to bends and torsions, and the rest as stiff, corresponding to bond stretches. For each group we report the relative eigenvalue error
\begin{equation}\label{eq:mre_definition}
    \text{MRE} = \underset{i \in \mathcal{S}}{\text{median}} \; \frac{\big| \hat{\omega}_i - \omega_i \big|}{\big| \omega_i \big|} ,
\end{equation}
with $\mathcal{S}$ the index set of the respective group. The median is used rather than the mean because soft modes have small $|\omega_i|$ by construction, so a few of them dominate the mean relative error without reflecting a correspondingly worse prediction. Finally, a generated structure is counted as a valid TS if its spectrum contains exactly one negative eigenvalue after projecting out the six translational and rotational modes.

\paragraph{Molecular conformer generation.}

For each test molecule with $L$ reference conformers, $K = 2L$ conformers are generated and compared against the reference ensemble. A conformer $C \in \mathbb{R}^{3N}$ assigns Cartesian coordinates to every atom of a given molecular graph, and the quality of an ensemble is assessed by how well the two sets cover one another. Following prior work, this is quantified by the coverage (COV) and the average minimum RMSD (AMR), each reported in a recall and a precision variant, with $\{C^*_l\}_{l=1}^{L}$ the reference conformers, $\{C_k\}_{k=1}^{K}$ the generated ones, and $\delta$ an RMSD threshold below which two conformers count as matching.

The recall variants measure how completely the reference ensemble is reproduced,
\begin{align}
    \text{COV-R} &= \frac{1}{L} \Big| \big\{ l \; \big| \; \exists k, \ \text{RMSD}(C^*_l, C_k) < \delta \big\} \Big| , \\
    \text{AMR-R} &= \frac{1}{L} \sum_{l=1}^{L} \min_{k} \ \text{RMSD}(C^*_l, C_k) ,
\end{align}
so that a high COV-R and a low AMR-R indicate that few reference conformers are missed. The precision variants exchange the roles of the two sets,
\begin{align}
    \text{COV-P} &= \frac{1}{K} \Big| \big\{ k \; \big| \; \exists l, \ \text{RMSD}(C_k, C^*_l) < \delta \big\} \Big| , \\
    \text{AMR-P} &= \frac{1}{K} \sum_{k=1}^{K} \min_{l} \ \text{RMSD}(C_k, C^*_l) ,
\end{align}
and measure instead how many generated conformers correspond to an actual reference conformer. Recall alone can be satisfied by an ensemble that is broader than the reference, and precision alone by one that is narrower, so the two are reported together.

We additionally compute for both, the generated and the reference ensemble of each molecule, the Boltzmann-weighted average of a property $A$,
\begin{equation}
    \langle A \rangle = \sum_{k} w_k A_k , \qquad
    w_k = \frac{\exp(-E_k / k_B T)}{\sum_{k'} \exp(-E_{k'} / k_B T)} ,
\end{equation}
with $E_k$ the electronic energy of conformer $k$ and $T = 298.15$~K. Energies, forces, dipole moments $\mu$ and HOMO-LUMO gaps $\Delta \epsilon$ are obtained with GFN2-xTB~\citep{bannwarth2019gfn2}. The reported metric is the absolute deviation between the two ensembles,
\begin{equation}
    \big| \Delta \langle A \rangle \big| = \big| \langle A \rangle_{\text{gen}} - \langle A \rangle_{\text{ref}} \big| ,
\end{equation}
averaged over the test molecules. The minimum energy $E_\text{min}$ is treated separately, since it characterizes the lowest-lying conformer rather than the ensemble as a whole and is therefore compared directly rather than through a weighted average. Each metric is evaluated at two stages. In the generated stage the properties are computed on the conformers as produced by the model, which reflects how far the ensemble sits from the PES. In the optimized stage they are computed after relaxing every conformer with GFN2-xTB, which removes local distortions and isolates the conformational composition of the ensemble from the accuracy of the individual geometries.

\section{Additional experiments}
\label{sec:additional_experiments}

\subsection{Cost of the curvature residual}
\label{sec:cost}

Imposing the residual at every training step is only practical because the surrogate is cheap to evaluate. A Hessian obtained by finite differences with GFN2-xTB takes on average $1.2$~s per structure on \swap{} and $4.1$~s on \tmc{}, the larger value reflecting the larger systems. HIP returns the same quantity in $51$ and $52$~ms, and its near-constant cost reflects the different scaling of the two approaches, since a finite-difference Hessian requires a number of force evaluations proportional to $3N$ while the predictor needs a single forward pass. The resulting overhead during training is modest. On \swap{}, the baseline requires $23$ hours on a single A100 GPU against $30$ hours for \method{}, an increase of $30\%$, with peak memory rising from $32$ to $36$~GB. Since the residual is confined to training, sampling remains unaffected.

\subsection{Transition state generation.}
\label{sec:additional_exp_ts}

\paragraph{Bond-resolved error analysis.}

Bonds are perceived with OpenBabel~\citep{o2011open} on the reference reactant and product structures, and their union defines the bond graph used throughout. Angles are taken as all triples of atoms connected through a common center, torsions as all chains of four connected atoms, and the non-bonded set as every atom pair not connected by a bond. For a set $\mathcal{S}$ of internal coordinates, the error of a generated structure $\hat{\mathbf{x}}_1$ against its reference TS $\mathbf{x}_1$ is
\begin{equation}
    \text{MAE}_{\mathcal{S}} = \frac{1}{|\mathcal{S}|} \sum_{s \in \mathcal{S}} \big| \hat{q}_s - q_s \big| ,
\end{equation}
where $q_s$ denotes the corresponding distance, angle, or torsion, and torsional differences are wrapped to $[0^\circ, 180^\circ]$. Table~\ref{tab:bond_metrics_summary} reports the mean and median of this quantity over the structures of each test set. On \swap{} and \tmc{} the median bond error decreases by $17$ and $13\%$ and the angle error by $8\%$ in both cases, while torsions and non-bonded distances change by at most $2\%$ and in some cases marginally for the worse.

\begin{table}[ht]
\centering
\caption{Structure error at the generated TSs resolved by internal coordinate. Mean and median of the per-structure mean absolute error. Best in bold, second-best underlined.}
\label{tab:bond_metrics_summary}
\begin{adjustbox}{max width=\linewidth}
\begin{tabular}{llcccccccc}
\toprule
 &  & \multicolumn{2}{c}{Bonded (\AA) $\downarrow$} & \multicolumn{2}{c}{Angle ($^\circ$) $\downarrow$} & \multicolumn{2}{c}{Torsion ($^\circ$) $\downarrow$} & \multicolumn{2}{c}{Non-bonded (\AA) $\downarrow$} \\
\cmidrule(lr){3-4} \cmidrule(lr){5-6} \cmidrule(lr){7-8} \cmidrule(lr){9-10}
Dataset & Model & Mean & Median & Mean & Median & Mean & Median & Mean & Median \\
\midrule
\tx & React-OT & \underline{0.024} & \underline{0.012} & \underline{2.46} & \textbf{1.50} & \textbf{8.94} & \textbf{4.64} & \textbf{0.077} & \textbf{0.047} \\
\tx & React-OT$_{\text{\method}}$ & \textbf{0.024} & \textbf{0.012} & \textbf{2.45} & \underline{1.53} & \underline{8.94} & \underline{4.65} & \underline{0.077} & \underline{0.047} \\
\midrule
\swap & React-OT & \underline{0.029} & \underline{0.018} & \underline{2.89} & \underline{1.99} & \textbf{8.01} & \textbf{5.03} & \textbf{0.082} & \underline{0.057} \\
\swap & React-OT$_{\text{\method}}$ & \textbf{0.027} & \textbf{0.015} & \textbf{2.78} & \textbf{1.84} & \underline{8.07} & \underline{5.06} & \underline{0.082} & \textbf{0.056} \\
\midrule
\tmc & React-OT & \underline{0.043} & \underline{0.032} & \underline{3.98} & \underline{3.29} & \underline{10.36} & \underline{7.79} & \underline{0.137} & \underline{0.112} \\
\tmc & React-OT$_{\text{\method}}$ & \textbf{0.039} & \textbf{0.028} & \textbf{3.80} & \textbf{3.04} & \textbf{10.26} & \textbf{7.70} & \textbf{0.136} & \textbf{0.111} \\
\bottomrule
\end{tabular}
\end{adjustbox}
\end{table}

\paragraph{Source of the reference Hessian.}

Table~\ref{tab:structural_metrics_summary_ablation_hessian} reports the two choices of reference Hessian, both trained on \swap{} under identical settings and compared against the purely structural baseline. With the surrogate on both sides, the median barrier error and maximum force decrease by $21$ and $21\%$. With a quantum-chemical reference they increase by $66$ and $48\%$ relative to the same baseline, so the residual is not merely uninformative in this case but actively harmful. The structural metrics are affected far less, rising by $2$ to $5\%$, which is consistent with the residual acting on the energetic rather than the geometric part of the objective.

\begin{table}[ht]
\centering
\caption{Structural and energetic metrics for TS generation. Best in bold, second-best underlined.}
\label{tab:structural_metrics_summary_ablation_hessian}
\begin{adjustbox}{max width=\linewidth}
\begin{tabular}{llcccccccc}
\toprule
 &  & \multicolumn{2}{c}{RMSD (\AA) $\downarrow$} & \multicolumn{2}{c}{D-MAE (\AA) $\downarrow$} & \multicolumn{2}{c}{$\Delta E$ (kcal/mol) $\downarrow$} & \multicolumn{2}{c}{$F_\text{max}$ (kcal/mol/\AA) $\downarrow$} \\
\cmidrule(lr){3-4} \cmidrule(lr){5-6} \cmidrule(lr){7-8} \cmidrule(lr){9-10}
Dataset & Model & Mean & Median & Mean & Median & Mean & Median & Mean & Median \\
\midrule
\swap & React-OT & \textbf{0.196} & \underline{0.134} & \underline{0.083} & \underline{0.060} & \underline{5.85} & \underline{2.53} & \underline{40.96} & \underline{33.50} \\
\swap & React-OT$_{\text{\method, HIP}}$ & \underline{0.197} & \textbf{0.133} & \textbf{0.082} & \textbf{0.058} & \textbf{5.19} & \textbf{2.00} & \textbf{32.69} & \textbf{26.43} \\
\swap & React-OT$_{\text{\method, QM}}$ & {0.199} & {0.137} & {0.087} & {0.062} & {7.30} & {4.20} & {56.34} & {49.70} \\
\bottomrule
\end{tabular}
\end{adjustbox}
\end{table}

The behavior in Table~\ref{tab:structural_metrics_summary_ablation_hessian} follows from how the surrogate error enters the two residuals. Writing $\epsilon(z) = H_{\mathrm{HIP}}(z) - H_{\mathrm{QM}}(z)$ for the surrogate error at a geometry $z$, the two choices decompose as
\begin{align}
  H_{\mathrm{HIP}}(\hat{x}_1) - H_{\mathrm{HIP}}(x_1)
    &= \underbrace{H_{\mathrm{QM}}(\hat{x}_1) - H_{\mathrm{QM}}(x_1)}_{\text{oracle residual}}
     + \underbrace{\epsilon(\hat{x}_1) - \epsilon(x_1)}_{\text{contamination}},
  \label{eq:hip-hip}\\[2pt]
  H_{\mathrm{HIP}}(\hat{x}_1) - H_{\mathrm{QM}}(x_1)
    &= \underbrace{H_{\mathrm{QM}}(\hat{x}_1) - H_{\mathrm{QM}}(x_1)}_{\text{oracle residual}}
     + \underbrace{\epsilon(\hat{x}_1)}_{\text{contamination}} .
  \label{eq:hip-qm}
\end{align}
The oracle term is what the residual should measure, namely the change in the true Hessian caused by the error in the endpoint estimate. The two choices differ only in what accompanies it. Eq.~\ref{eq:hip-qm} carries the full surrogate error, which does not vanish as $\hat{x}_1 \to x_1$, whereas Eq.~\ref{eq:hip-hip} carries only the difference of the surrogate errors at two nearby geometries, which vanishes whenever $\epsilon$ varies smoothly in $z$, however large $\epsilon$ itself may be.

\begin{figure}[ht]
\centering
\includegraphics[width=0.5\textwidth]{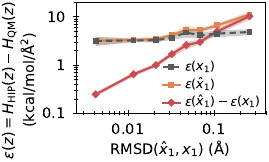}
\caption{\textbf{Surrogate error cancels between the two endpoints.} Mean absolute surrogate error $\epsilon(z) = H_{\mathrm{HIP}}(z) - H_{\mathrm{QM}}(z)$ at the reference TS $x_1$ and at the endpoint estimate $\hat{x}_1$, together with their difference, binned by the RMSD between the two geometries on \swap{}. Markers joined by lines are binned means and shaded regions are confidence intervals of the mean.}
\label{fig:error_cancellation}
\end{figure}

To verify this, we used the one-shot predictions $\hat{x}_1$ on reactions from \swap shown in Figure~\ref{fig:residual_source} and compute the Hessian using the GFN2-xTB level of theory. Figure~\ref{fig:error_cancellation} shows $\langle|\epsilon(x_1)|\rangle$, $\langle|\epsilon(\hat{x}_1)|\rangle$ and $\langle|\epsilon(\hat{x}_1) - \epsilon(x_1)|\rangle$ against the RMSD between the two geometries. Below $0.01$~\AA{} the two individual errors are nearly equal, at $3.16$ and $3.18$~kcal/mol/\AA$^2$, while their difference is $0.25$, which is $8\%$ of the individual error, so $92\%$ of the surrogate error cancels in Eq.~\ref{eq:hip-hip}.

\paragraph{Bias from the surrogate.}

Since \method{} is trained against HIP predictions, the evaluation could in principle reward structures on which the surrogate agrees with itself rather than structures that are physically correct. To test this, we repeat the Hessian evaluation of Table~\ref{tab:hessian_metrics_summary} using HIP in place of quantum chemistry for both the generated and the reference structures. The results are reported in Table~\ref{tab:hessian_metrics_summary_hip}. Both error measures improve under \method{} on all three datasets, and the valid TS rate rises on each, in agreement with the quantum-chemical evaluation. The absolute values differ, most visibly in the valid TS rate on \tmc{}, where HIP classifies a larger fraction of structures as valid than quantum chemistry does, but the ordering of the two models is the same throughout.

\begin{table}[ht]
\centering
\caption{Hessian metrics for TS generation. Hessians of both the generated and the ground-truth structures are computed with the HIP model of the respective dataset. Best in bold, second-best underlined.}
\label{tab:hessian_metrics_summary_hip}
\begin{adjustbox}{max width=\linewidth}
\begin{tabular}{llccccc}
\toprule
 &  & \multicolumn{2}{c}{Hessian MAE (kcal/mol/\AA$^2$) $\downarrow$} & \multicolumn{2}{c}{Eigenvalue MAE (kcal/mol/\AA$^2$) $\downarrow$} & \multicolumn{1}{c}{Valid TS (\%) $\uparrow$} \\
\cmidrule(lr){3-4} \cmidrule(lr){5-6} \cmidrule(lr){7-7}
Dataset & Model & Mean & Median & Mean & Median & Mean \\
\midrule
\tx & React-OT & \underline{13.26} & \textbf{7.78} & \underline{25.36} & \underline{17.61} & \underline{71.00} \\
\tx & React-OT$_{\text{\method}}$ & \textbf{13.20} & \underline{7.93} & \textbf{24.86} & \textbf{16.94} & \textbf{74.00} \\
\midrule
\swap & React-OT & \underline{11.72} & \underline{8.48} & \underline{28.12} & \underline{21.71} & \underline{61.00} \\
\swap & React-OT$_{\text{\method}}$ & \textbf{11.04} & \textbf{7.49} & \textbf{23.51} & \textbf{17.08} & \textbf{64.00} \\
\midrule
\tmc & React-OT & \underline{11.19} & \underline{9.35} & \underline{36.32} & \underline{29.37} & \underline{49.00} \\
\tmc & React-OT$_{\text{\method}}$ & \textbf{10.38} & \textbf{8.56} & \textbf{28.59} & \textbf{22.35} & \textbf{55.00} \\
\bottomrule
\end{tabular}
\end{adjustbox}
\end{table}

\paragraph{Spectrally resolved Hessian metrics.}

Table~\ref{tab:hessian_metrics_summary_detailed} resolves the Hessian error by mode, separating soft from stiff vibrations at a reference wavenumber of $500$~cm$^{-1}$ and reporting the overlap of the reactive mode separately. On \swap{} and \tmc{} both groups improve under \method{}, the soft modes by $19$ and $14\%$ and the stiff modes by $27$ and $28\%$ in the median relative error. On \tx{} the improvement is smaller in both groups. The reactive mode changes little on any dataset, with the median overlap rising by less than one point and the mean on \tx{} marginally lower, so the structural objective already reproduces this direction well and the residual acts mainly on the remaining vibrational modes.

\begin{table}[ht]
\centering
\caption{Hessian metrics for TS generation, computed against ground-truth QM Hessians. Soft modes are assigned by $|\lambda_i| < 500$ cm$^{-1}$. Best in bold, second-best underlined.}
\label{tab:hessian_metrics_summary_detailed}
\begin{adjustbox}{max width=\linewidth}
\begin{tabular}{llcccccc}
\toprule
 &  & \multicolumn{2}{c}{Soft Eigenvalue MRE (\%) $\downarrow$} & \multicolumn{2}{c}{Stiff Eigenvalue MRE (\%) $\downarrow$} & \multicolumn{2}{c}{Neg. Mode Overlap (\%) $\uparrow$} \\
\cmidrule(lr){3-4} \cmidrule(lr){5-6} \cmidrule(lr){7-8}
Dataset & Model & Mean & Median & Mean & Median & Mean & Median \\
\midrule
\tx & React-OT & \textbf{22.57} & \underline{11.78} & \underline{2.53} & \underline{1.80} & \textbf{92.35} & \underline{99.05} \\
\tx & React-OT$_{\text{\method}}$ & \underline{22.60} & \textbf{10.44} & \textbf{2.45} & \textbf{1.68} & \underline{92.04} & \textbf{99.08} \\
\midrule
\swap & React-OT & \underline{25.04} & \underline{16.93} & \underline{3.48} & \underline{2.61} & \underline{82.42} & \underline{95.92} \\
\swap & React-OT$_{\text{\method}}$ & \textbf{21.89} & \textbf{13.79} & \textbf{2.68} & \textbf{1.90} & \textbf{85.29} & \textbf{96.57} \\
\midrule
\tmc & React-OT & \underline{22.27} & \underline{17.30} & \underline{4.39} & \underline{3.60} & \underline{80.91} & \underline{94.84} \\
\tmc & React-OT$_{\text{\method}}$ & \textbf{19.23} & \textbf{14.89} & \textbf{3.25} & \textbf{2.59} & \textbf{82.91} & \textbf{95.41} \\
\bottomrule
\end{tabular}
\end{adjustbox}
\end{table}

\paragraph{Downstream behavior.}

Beyond the static metrics, we measure downstream utility by running a saddle-point optimization from each generated structure at the level of theory of the respective dataset, which is $\omega$B97X/6-31G(d) for \tx{} and GFN2-xTB for \swap{} and \tmc{}. The difference of two orders of magnitude in the optimization time reported in Table~\ref{tab:optimization_metrics_summary} reflects this difference in level of theory rather than a property of the models. The median RMSD after optimization is zero for both variants on all three datasets, so the optimizations recover the reference TS exactly for the majority of reactions. The nonzero means, between $0.08$ and $0.27$~\AA{}, come from a minority of cases that converge to a different saddle point, and their share is the same for both variants. The comparison therefore concerns how quickly the intended structure is reached rather than whether it is reached. On \swap{} the mean number of steps decreases by $18\%$ and the mean optimization time by $23\%$. On \tx{} the effect is small, with both the mean and median number of steps decreasing slightly. On \tmc{} the median decreases while the mean rises, so a minority of difficult cases requires more iterations under \method{}. The mean optimization time nevertheless decreases on this dataset as well, which indicates that the individual steps are cheaper, as expected for structures that are closer to the physical geometry and for which the SCF converges more readily.

\begin{table}[ht]
\centering
\caption{Optimization metrics for TS generation comparing post-optimization TS structures. Best in bold, second-best underlined.}
\label{tab:optimization_metrics_summary}
\begin{adjustbox}{max width=\linewidth}
\begin{tabular}{llccccccc}
\toprule
 &  & \multicolumn{2}{c}{RMSD (\AA) $\downarrow$} & \multicolumn{2}{c}{D-MAE (\AA) $\downarrow$} & \multicolumn{2}{c}{Opt. Steps} & \multicolumn{1}{c}{Opt. Time (s)} \\
\cmidrule(lr){3-4} \cmidrule(lr){5-6} \cmidrule(lr){7-8} \cmidrule(lr){9-9}
Dataset & Model & Mean & Median & Mean & Median & Mean & Median & Mean \\
\midrule
\tx & React-OT & \textbf{0.080} & 0.000 & 0.020 & 0.000 & \underline{44.63} & \underline{39.00} & \underline{1505.50} \\
\tx & React-OT$_{\text{\method}}$ & \underline{0.090} & 0.000 & 0.020 & 0.000 & \textbf{44.16} & \textbf{37.00} & \textbf{1500.70} \\
\midrule
\swap & React-OT & 0.120 & 0.000 & 0.060 & 0.000 & \underline{55.70} & \underline{40.00} & \underline{8.54} \\
\swap & React-OT$_{\text{\method}}$ & 0.120 & 0.000 & 0.060 & 0.000 & \textbf{45.98} & \textbf{39.50} & \textbf{6.54} \\
\midrule
\tmc & React-OT & \underline{0.270} & 0.000 & \underline{0.100} & 0.000 & \textbf{116.11} & \underline{79.00} & \underline{20.74} \\
\tmc & React-OT$_{\text{\method}}$ & \textbf{0.260} & 0.000 & \textbf{0.090} & 0.000 & \underline{121.37} & \textbf{78.00} & \textbf{20.45} \\
\bottomrule
\end{tabular}
\end{adjustbox}
\end{table}

\paragraph{Accuracy of the endpoint estimate.}

Finally, we examine the effect of the endpoint estimate itself. Unrolling two ODE steps rather than one brings $\hat{x}_1$ closer to the true endpoint and thereby narrows Jensen's gap, at the cost of a second model call per training step. Table~\ref{tab:structural_metrics_summary_ablation_integration} shows that this yields a further improvement of roughly $11\%$ in both the median barrier error and the median maximum force, reducing them to $1.78$~kcal/mol and $23.46$~kcal/mol/\AA{}, while the global structural metrics again remain unchanged. The improvement comes at a substantial price, with training time rising by $57\%$ to $47$ hours and a corresponding increase in memory. We therefore use a single step throughout, and report this variant as an option where maximal physical fidelity outweighs the cost of training.

\begin{table}[ht]
\centering
\caption{Structural and energetic metrics for TS generation. Best in bold, second-best underlined.}
\label{tab:structural_metrics_summary_ablation_integration}
\begin{adjustbox}{max width=\linewidth}
\begin{tabular}{llcccccccc}
\toprule
 &  & \multicolumn{2}{c}{RMSD (\AA) $\downarrow$} & \multicolumn{2}{c}{D-MAE (\AA) $\downarrow$} & \multicolumn{2}{c}{$\Delta E$ (kcal/mol) $\downarrow$} & \multicolumn{2}{c}{$F_\text{max}$ (kcal/mol/\AA) $\downarrow$} \\
\cmidrule(lr){3-4} \cmidrule(lr){5-6} \cmidrule(lr){7-8} \cmidrule(lr){9-10}
Dataset & Model & Mean & Median & Mean & Median & Mean & Median & Mean & Median \\
\midrule
\swap & React-OT & 0.196 & \underline{0.134} & 0.083 & 0.060 & 5.85 & 2.53 & 40.96 & 33.50 \\
\swap & React-OT$_{\text{\method}}$ & 0.197 & \textbf{0.133} & \underline{0.082} & \underline{0.058} & \underline{5.19} & \underline{2.00} & \underline{32.69} & \underline{26.43} \\
\swap & React-OT$_{\text{\method, nfe=2}}$ & 0.196 & 0.136 & \textbf{0.081} & \textbf{0.057} & \textbf{4.84} & \textbf{1.78} & \textbf{28.40} & \textbf{23.46} \\
\bottomrule
\end{tabular}
\end{adjustbox}
\end{table}

\paragraph{Formulation of the curvature residual.}

The residual of Eq.~\ref{eq:residual_mae} constrains the full Hessian, treating all directions of the spectrum alike. Since the type of stationary point is determined by its lowest modes, a residual restricted to that part of the spectrum could in principle deliver the same information at a fraction of the constraint, which is what the subspace variant of Eq.~\ref{eq:residual_sub} tests. We set $k=8$ so that the two lowest vibrational modes are covered, since the six rigid-body modes are retained rather than projected out. Following~\citet{burger2025shoot}, we avoid the projection because it introduces numerical instabilities in the backward pass, and the same choice applies to the entrywise residual, which is evaluated on the unprojected matrix throughout. Tables~\ref{tab:structural_metrics_summary_ablation_loss} and~\ref{tab:hessian_metrics_summary_ablation_loss} report the three formulations on \swap{}, together with the purely structural baseline.

The subspace residual improves on the baseline by $6\%$ in both the median barrier error and the median maximum force, and raises the valid TS rate by $1\%$. The full residual improves the same quantities by $21\%$ and the valid TS rate by $10\%$ The difference is largest among the stiff modes, where the median relative eigenvalue error falls by $7\%$ under the subspace variant and by $27\%$ under the full one, and it is smallest for the reactive mode, whose overlap changes by less than one point under either formulation.

The combination of both terms behaves almost identically to the full residual. The subspace term therefore adds nothing once the full matrix is constrained, and the small consistent degradation suggests that it competes with the denser signal rather than complementing it.

\begin{table}[ht]
\centering
\caption{Structural and energetic metrics for TS generation. Best in bold, second-best underlined.}
\label{tab:structural_metrics_summary_ablation_loss}
\begin{adjustbox}{max width=\linewidth}
\begin{tabular}{llcccccccc}
\toprule
 &  & \multicolumn{2}{c}{RMSD (\AA) $\downarrow$} & \multicolumn{2}{c}{D-MAE (\AA) $\downarrow$} & \multicolumn{2}{c}{$\Delta E$ (kcal/mol) $\downarrow$} & \multicolumn{2}{c}{$F_\text{max}$ (kcal/mol/\AA) $\downarrow$} \\
\cmidrule(lr){3-4} \cmidrule(lr){5-6} \cmidrule(lr){7-8} \cmidrule(lr){9-10}
Dataset & Model & Mean & Median & Mean & Median & Mean & Median & Mean & Median \\
\midrule
\swap & React-OT & \underline{0.196} & 0.134 & 0.083 & 0.060 & 5.85 & 2.53 & 40.96 & 33.50 \\
\swap & React-OT$_{\text{\method, full}}$ & 0.197 & 0.133 & 0.082 & 0.058 & \textbf{5.19} & \textbf{2.00} & \underline{32.69} & \textbf{26.43} \\
\swap & React-OT$_{\text{\method, subspace}}$ & \textbf{0.194} & 0.133 & 0.082 & 0.058 & 5.62 & 2.38 & 38.88 & 31.64 \\
\swap & React-OT$_{\text{\method, combined}}$ & 0.199 & 0.141 & 0.083 & 0.060 & \underline{5.39} & \underline{2.01} & \textbf{32.60} & \underline{26.89} \\
\bottomrule
\end{tabular}
\end{adjustbox}
\end{table}

\begin{table}[ht]
\centering
\caption{Hessian metrics for TS generation. Hessians of both the generated and the ground-truth structures are computed with quantum chemistry at the level of theory of the respective dataset. Soft modes are assigned by $|\lambda_i| < 500$ cm$^{-1}$. Best in bold, second-best underlined.}
\label{tab:hessian_metrics_summary_ablation_loss}
\begin{adjustbox}{max width=\linewidth}
\begin{tabular}{llccccccc}
\toprule
 &  & \multicolumn{2}{c}{Soft Eigenvalue MRE (\%) $\downarrow$} & \multicolumn{2}{c}{Stiff Eigenvalue MRE (\%) $\downarrow$} & \multicolumn{2}{c}{Neg. Mode Overlap (\%) $\uparrow$} & \multicolumn{1}{c}{Valid TS (\%) $\uparrow$} \\
\cmidrule(lr){3-4} \cmidrule(lr){5-6} \cmidrule(lr){7-8} \cmidrule(lr){9-9}
Dataset & Model & Mean & Median & Mean & Median & Mean & Median & Mean \\
\midrule
\swap & React-OT & 25.04 & 16.93 & 3.48 & 2.61 & 82.42 & 95.92 & 57.00 \\
\swap & React-OT$_{\text{\method, full}}$ & \textbf{21.89} & \textbf{13.79} & \textbf{2.68} & \textbf{1.90} & \textbf{85.29} & \textbf{96.57} & \textbf{67.00} \\
\swap & React-OT$_{\text{\method, subspace}}$ & 24.76 & 15.49 & 3.24 & 2.44 & 83.42 & 96.12 & 58.00 \\
\swap & React-OT$_{\text{\method, combined}}$ & \underline{23.16} & \underline{14.12} & \underline{2.73} & \underline{1.91} & \underline{84.26} & \underline{96.26} & \underline{66.00} \\
\bottomrule
\end{tabular}
\end{adjustbox}
\end{table}

\paragraph{Order of the physical signal.}

To test whether curvature is required or whether a first-order signal would suffice, we replace the Hessian in Eq.~\ref{eq:general_residual} by the forces predicted at the two endpoints. Writing $F_\phi : \mathbb{R}^{3N \times 3} \to \mathbb{R}^{3N \times 3}$ for the forces returned by the same frozen surrogate, the residual becomes
\begin{equation}
    \mathcal{R}_{\mathrm{force}}(\hat{x}_1, x_1) = \frac{1}{3N} \sum_{i,j} \big| \big[F_\phi(\hat{x}_1)\big]_{ij} - \big[F_\phi(x_1)\big]_{ij} \big| ,
    \label{eq:residual_force}
\end{equation}
so that both sides are again evaluated with the surrogate and its systematic error cancels as before. Since forces and curvature differ in units and magnitude, the weight $\lambda$ and the exponent $p$ are optimized separately for this variant on the validation set, using the same procedure as for the curvature residual. The results on \swap{} are reported in Table~\ref{tab:structural_metrics_summary_ablation_force}. The force residual does improve on the purely structural baseline, reducing the median barrier error and maximum force by $7\%$ each, but the curvature residual reduces both by $21\%$. The first-order signal thus recovers about a third of the available gain, and does so even for the maximum force, the quantity it supervises directly.

\begin{table}[ht]
\centering
\caption{Structural and energetic metrics for TS generation. Best in bold, second-best underlined.}
\label{tab:structural_metrics_summary_ablation_force}
\begin{adjustbox}{max width=\linewidth}
\begin{tabular}{llcccccccc}
\toprule
 &  & \multicolumn{2}{c}{RMSD (\AA) $\downarrow$} & \multicolumn{2}{c}{D-MAE (\AA) $\downarrow$} & \multicolumn{2}{c}{$\Delta E$ (kcal/mol) $\downarrow$} & \multicolumn{2}{c}{$F_\text{max}$ (kcal/mol/\AA) $\downarrow$} \\
\cmidrule(lr){3-4} \cmidrule(lr){5-6} \cmidrule(lr){7-8} \cmidrule(lr){9-10}
Dataset & Model & Mean & Median & Mean & Median & Mean & Median & Mean & Median \\
\midrule
\swap & React-OT & \underline{0.196} & 0.134 & 0.083 & 0.060 & 5.85 & 2.53 & 40.96 & 33.50 \\
\swap & React-OT$_{\text{\method, force}}$ & \textbf{0.195} & \textbf{0.132} & 0.082 & 0.058 & \underline{5.62} & \underline{2.35} & \underline{37.91} & \underline{31.07} \\
\swap & React-OT$_{\text{\method, Hessian}}$ & 0.197 & \underline{0.133} & 0.082 & 0.058 & \textbf{5.19} & \textbf{2.00} & \textbf{32.69} & \textbf{26.43} \\
\bottomrule
\end{tabular}
\end{adjustbox}
\end{table}

\subsection{Molecular conformer generation.}
\label{sec:additional_exp_conformer}

\paragraph{Validity and optimization of generated conformers.}

Curvature is assessed directly by computing Hessians with GFN2-xTB for up to 32 conformers per molecule and testing whether each is a local minimum, that is, whether its spectrum is free of negative eigenvalues. Table~\ref{tab:etflow_hessian_metrics_summary} reports the resulting fraction of valid conformers per molecule. The reference ensembles of GEOM-QM9 reach $95.7\%$ on average rather than $100\%$, although they were generated with the same semi-empirical method, which we attribute to numerical precision and differences in code version between the original CREST run and our evaluation. This value sets the practical ceiling against which the generated ensembles are measured. \method{} raises the median from $96.5$ to $100\%$, so the majority of molecules yield ensembles in which every conformer is a local minimum, while the mean is marginally lower at $87.6$ against $89.0\%$.

Table~\ref{tab:etflow_optimization_metrics_summary} reports the number of GFN2-xTB optimization steps required to relax each generated conformer to a local minimum. \method{} reduces this from $5.4$ to $4.8$ in the median and from $6.1$ to $5.8$ in the mean. Since every step corresponds to an energy and force evaluation, the reduction translates directly into a lower cost for any downstream calculation that begins from the generated ensemble.

\begin{table}[ht]
\centering
\caption{Validity rate of reference and generated conformers as local Hessian minima on GEOM-QM9. Best in bold, second-best underlined.}
\label{tab:etflow_hessian_metrics_summary}
\begin{adjustbox}{max width=\linewidth}
\begin{tabular}{lccc}
\toprule
 & \multicolumn{2}{c}{Gen. Valid Equilibrium (\%) $\uparrow$} \\
\cmidrule(lr){2-3}
Model & Mean & Median \\
\midrule
ET-Flow & \textbf{89.0} & \underline{96.5} \\
ET-Flow$_{\text{\method}}$ & \underline{87.6} & \textbf{100.0} \\
\bottomrule
\end{tabular}
\end{adjustbox}
\end{table}

\begin{table}[ht]
\centering
\caption{Optimization outcome for generated conformers on GEOM-QM9. Best in bold, second-best underlined.}
\label{tab:etflow_optimization_metrics_summary}
\begin{adjustbox}{max width=\linewidth}
\begin{tabular}{lcc}
\toprule
 & \multicolumn{2}{c}{Opt. Steps} \\
\cmidrule(lr){2-3}
Model & Mean & Median \\
\midrule
ET-Flow & \underline{6.1} & \underline{5.4} \\
ET-Flow$_{\text{\method}}$ & \textbf{5.8} & \textbf{4.8} \\
\bottomrule
\end{tabular}
\end{adjustbox}
\end{table}

\paragraph{Comparison against a stronger structural baseline.}

Tables~\ref{tab:etflow_structural_metrics_summary_ablation_ditmc} and~\ref{tab:etflow_ensemble_metrics_summary_ablation_ditmc} compare ET-Flow trained with \method{} against DiTMC~\citep{frank2025sampling}, reproduced from the public checkpoint. DiTMC is the stronger generative model, with a lower median AMR in recall and precision. Despite this, the two are comparable on the ensemble properties, where DiTMC leads by $25\%$ in the median energy deviation and $20\%$ in the minimum energy, and the dipole moment is identical. On the maximum force the ordering reverses, with \method{} reducing the median deviation by $17\%$ and the mean by $22\%$. Ensemble properties are averages over the generated conformers and therefore reflect how completely the reference ensemble is recovered, whereas the maximum force is a property of each individual structure. The purpose of this comparison is to place the size of the effect in context rather than to rank the two models, which differ in architecture, training data, and budget. Curvature supervision applied to the weaker generative model is sufficient to reach comparable physical fidelity, and to surpass it where the metric is sensitive to individual structures. \method{} is architecture-agnostic, as the residual is defined on the endpoint estimate and not on the model that produces it, so it can be combined with DiTMC or any other flow matching model.

\begin{table}[ht]
\centering
\caption{Molecular conformer generation. results on GEOM-QM9 ($\delta$ = 0.5\AA). Best in bold, second-best underlined.}
\label{tab:etflow_structural_metrics_summary_ablation_ditmc}
\begin{adjustbox}{max width=\linewidth}
\begin{tabular}{lcccccccc}
\toprule
 & \multicolumn{2}{c}{Coverage-R (\%) $\uparrow$} & \multicolumn{2}{c}{AMR-R (\AA) $\downarrow$} & \multicolumn{2}{c}{Coverage-P (\%) $\uparrow$} & \multicolumn{2}{c}{AMR-P (\AA) $\downarrow$} \\
\cmidrule(lr){2-3} \cmidrule(lr){4-5} \cmidrule(lr){6-7} \cmidrule(lr){8-9}
Model & Mean & Median & Mean & Median & Mean & Median & Mean & Median \\
\midrule
DiTMC+PE(3)-B$_\text{reproduced}$ & \underline{95.02} & 100.00 & \textbf{0.075} & \textbf{0.023} & \textbf{93.29} & 100.00 & \textbf{0.090} & \textbf{0.032} \\
ET-Flow$_\text{\method}$  & \textbf{95.97} & 100.00 & \underline{0.080} & \underline{0.035} & \underline{91.33} & 100.00 & \underline{0.121} & \underline{0.054} \\
\bottomrule
\end{tabular}
\end{adjustbox}

\vspace{0.8em}

\caption{Ensemble property deviations (mean absolute difference vs. reference) on GEOM-QM9, before geometry optimization. Best in bold, second-best underlined.}
\label{tab:etflow_ensemble_metrics_summary_ablation_ditmc}
\begin{adjustbox}{max width=\linewidth}
\begin{tabular}{llcccccccccc}
\toprule
 &  & \multicolumn{2}{c}{$|\Delta \langle E \rangle|$ (kcal/mol)} $\downarrow$ & \multicolumn{2}{c}{$|\Delta \langle F_\text{max} \rangle|$ (kcal/mol/Å)} $\downarrow$ & \multicolumn{2}{c}{$|\Delta \langle \mu \rangle|$ (D)} $\downarrow$ & \multicolumn{2}{c}{$|\Delta \langle \Delta \epsilon \rangle|$ (kcal/mol)} $\downarrow$ & \multicolumn{2}{c}{$|\Delta \langle E_{min} \rangle| $ (kcal/mol)} $\downarrow$ \\
\cmidrule(lr){3-4} \cmidrule(lr){5-6} \cmidrule(lr){7-8} \cmidrule(lr){9-10} \cmidrule(lr){11-12}
Model & Stage & Mean & Median & Mean & Median & Mean & Median & Mean & Median & Mean & Median \\
\midrule
DiTMC+PE(3)-B$_\text{reproduced}$ & Generated & \textbf{0.27} & \textbf{0.12} & \underline{8.14} & \underline{6.96} & {0.15} & 0.04 & {1.05} & \underline{0.57} & \textbf{0.25} & \textbf{0.10} \\
ET-Flow$_\text{\method}$ & Generated & \underline{0.33} & \underline{0.15} & \textbf{6.37} & \textbf{5.81} & {0.15} & {0.04} & {1.05} & \textbf{0.44} & \underline{0.31} & \underline{0.12} \\
\bottomrule
\end{tabular}
\end{adjustbox}
\end{table}